# File System Design Approaches

Dr. Brijender Kahanwal
Department of Computer Science & Engineering
Galaxy Global Group of Institutions
Dinarpur, Ambala, Haryana, INDIA
imkahanwal@gmail.com

*Abstract*—**In this article, the file system development design approaches are discussed. The selection of the file system design approach is done according to the needs of the developers what are the needed requirements and specifications for the new design. It allowed us to identify where our proposal fitted in with relation to current and past file system development. Our experience with file system development is limited so the research served to identify the different techniques that can be used. The variety of file systems encountered show what an active area of research file system development is. The file systems may be from one of the two fundamental categories. In one category, the file system is developed in user space and runs as a user process. Another file system may be developed in the kernel space and runs as a privileged process. Another one is the mixed approach in which we can take the advantages of both aforesaid approaches. Each development option has its own pros and cons.  In this article, these design approaches are discussed.**

## I. INTRODUCTION

The file system is a mature enough abstraction that it should be implemented inside the operating system kernel for optimal performance. Data storage and high-level naming are fundamental features of modern computer systems, along with process and communication abstractions. The file systems are oriented towards access to slow devices such as disks and terminals. In these cases, it is more natural to use read and write procedures that move data from the slower devices into main memory. Open and close procedures provide convenient points at which the system can set up state about the underlying device in order to optimize later read and write calls. In particular, setup can be performed to optimize network accesses. Sharing resources is also easier via the file system interface because the calls into the file system interface provide well defined points at which to check for consistency among shared copies.

Accesses to file system data are by and large sequential. So, the systems that layer the file system on top of mapped files often suffer performance problems. The resources controlled by the file system are ultimately accessed via kernel-resident device drivers.  A user process can implement any semantics it chooses for a pseudo-device or a pseudo-file-system.

The experience with file system development is limited so the research served to identify the different techniques that can be used. The variety of file systems encountered show what an active area of research file system development is. The file systems researched fell into one of the following four categories:
1. The file system is developed in user space and runs as a user process.
2. The file system is developed in the user space using FUSE (File system in USEr space) kernel module and runs as a user process.
3. The file system is developed in the kernel and runs as a privileged process. The file system development implements all of the file system functionality.
4. The new functionality that the file system provides is stacked on top of existing file system functionality using stackable layers.

Every aforesaid technique is explored one by one with their peculiarities and advantages in the following subsections of this article.

## II. USER PROCESS FILE SYSTEM

The idea of developing a file system as a user process is appealing for a variety of reasons not least of which being that it is simpler than other techniques. By developing the file system as a user level process, the complexity of kernel level programming can be avoided. This simplifies the development process enormously, as developing in the kernel is more restrictive than user level development. The standard development, debugging tools and programming libraries can be used. This helps to reduce the time required to implement the file system.

One of the most advantages of developing a file system as a user level process is that the file system can be installed by a user without the assistance of a system administrator. This provides the user with greater flexibility in how they use files. Fig. 1illustrates how a file system developed to run in user space interacts with the local and remote operating systems.

A user process requests access to a file from a user-space file system. The request is routed through the kernel. The steps in the communication show how a request by a user process results in a context switch in to and out of the kernel.

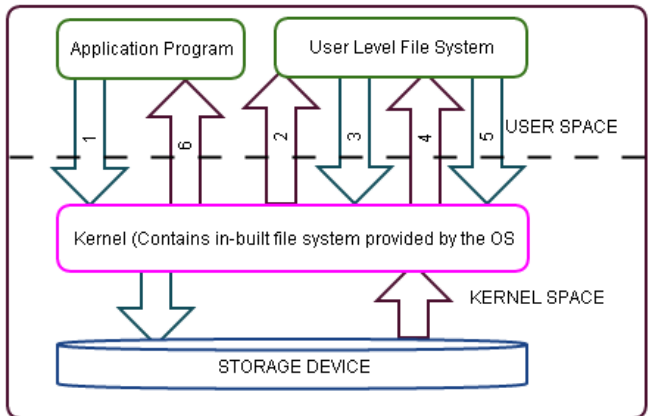

Figure 1. File System Implemented as a User Level Process.

a) The user makes a request to read a file. This results in a call to the kernel that forwards the call on to the user level file system,
b) The user level file system makes another call to the kernel,
c) To retrieve the data required by the read from the storage media. The kernel passes the data back to the user level file system at,
d) The user level file system now calls the kernel again to pass the data back to the user process,
e) The kernel completes the read command by delivering the data to the user process at,
f) This results in two additional context switches to the kernel than a normal read.

There are many examples in the literature of file systems developed as user processes; these include CFS and TCFS. All of the file systems that are implemented as a user level process are susceptible to a major performance problem. The use of a user level process requires additional context switches that increase the overhead of every system call and thus reduce performance.

### III. USER PROCESS FILE SYSTEM USING FUSE KERNEL MODULE

FUSE is a recent example of a general trend in operating system design: reducing the complexity of the kernel. FUSE consists of a kernel module and a user space library that allow a non-privileged user to create a file system without modifying the operating system kernel. The module provides a relatively simple API, and acts as a translator to the actual kernel interface.

The API uses function callbacks foremost common file system functions (read (), write (), close (), read (), write (), opendir (), readdir (), mkdir (), open (), create (), etc.). A file system developer implements his or her own versions of these functions and registers them with the FUSE module. A major benefit of using FUSE is the API simplicity, which makes file system development quick and relatively easy, and the fact that the file system runs in user space, which makes it possible to use the usual debugging tools.

Fig. 2 shows the path of a call to a FUSE file system. The FUSE kernel module receives a request to a file system, and passes the request on to libfuse which generates a call to the file system. This file system is then free to do whatever any normal user space program can do, e.g. use SSH to interface with a file system at a remote location.

This is another approach for development of file system as a user level process; the complexity of kernel level programming can be avoided. This is more simplified way of the development. In this the FUSE kernel level module is used which is a complete Application Programming Interface (API). This module is available for the most of the standard languages like C, C++, Java, Perl etc. The developer can choose the language of his comfort. This helps to reduce the time required to implement the file system.

It also has the same advantage as one of the previous approach that the file system can be installed by a user without the assistance of a system administrator. This provides the user with greater flexibility in how they use files. Fig. 2 illustrates how a file system developed to run in user space interacts with the local and remote operating systems.

The working of the FUSE file system is as follows:
a) File system request, by the application to the kernel
b) VFS issues the request and send it to FUSE Module within the kernel
c) FUSE module receive it and forward it to the FUSE file system
d) The FUSE file system receives it and gives response to the FUSE module back
e) FUSE module receives the response from the file system and send it to the VFS kernel module
f) Finally the response is given to the concern Application

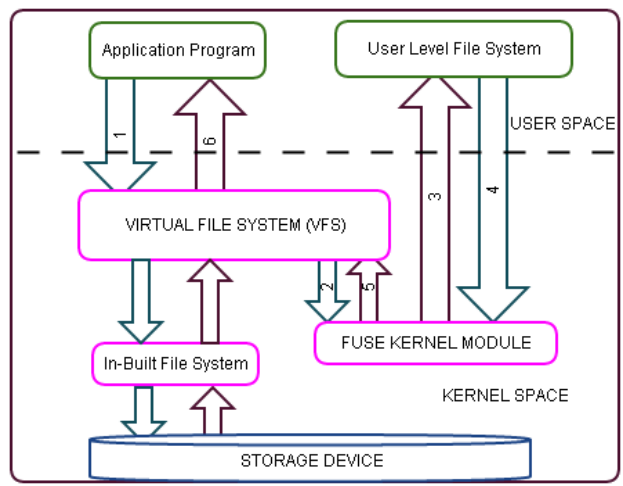

Figure 2. The path of a file system call, and the FUSE module.

The examples of such type of development are EncFS, encfs, cvsfs, sshfs, ZFS; AAFS, Direct Access File System (DAFS), developed by Magoutis et al. is an example of file system implemented in a user-space.

It is a framework of implementing file systems in user space. It eliminates the need to write kernel modules in order to implement a file system. It was firstly, originated from the Linux developers community. It has basically two parts in it one of them is a kernel module and another one is user space library (libfuse). The user space library provides a framework and Application Programming Interface (API). The API helps us to implement a file system completely within user space. The API uses function callbacks foremost common file system functions (open (), create (), read (), write (), unlink (), etc.). A file system developer implements his or her own versions of these functions and registers them with the FUSE module.

The FUSE library (libfuse) interacts with a kernel module. The FUSE kernel module implements a virtual file system interface and a character device interface. The kernel module and the user space library communicate through the character device. These two portions collaborate with each other through a queue mechanism to organize the file system calls to FUSE operations callback via Fuse library.

The Fuse library opens a character device and goes into a loop, (a) Read Device, (b) Process (unorganized IN parameters; call the Fuse operation callback; organize back the result) (c) Write Device.

Developing in-kernel file systems is a challenging task, because of many reasons. These are as follows:
(a.) The kernel code lacks memory protection,
(b.) It requires great attention to use of synchronization primitives,
(c.) These can be written in C language only,
(d.) Debugging is a tedious task,
(e.) Errors in the developed file systems can require rebooting the system,
(f.) Porting of the kernel file systems requires significant changes in the design and implementation, and
(g.) These can be mounted only with super-user privileges.

By developing of file systems in user space eliminates all the above issues. At the same time, the research in the area of storage and file systems increasing involves the addition of rich functionalities over the underlying, as opposed to designing the low-level file systems directly. On the other end, by developing in user space, the programmer has a wide range of programming languages, third-party tools and libraries. The file system should be highly portable to other operating systems. The kernel remains smaller and more reliable. Because of these, we have developed the Java File Security System (JFSS). As the name describes, Java programming language has been selected for the development which is well known for the feature of high portability. There are so many advantages to develop the file systems using FUSE, which are as follows:

(a.) We can implement the fully functional file system in user space,
(b.) FUSE provides file systems in user space
(c.) Quick file system development
(d.) File systems that need user-level services
(e.) Extending existing file systems
(f.) Very efficient user & kernel space interface,
(g.) File system developer can use all the POSIX and user space libraries,
(h.) Completely isolated development from the kernel programming & operating system intricacies,
(i.) Many language bindings: C, C++, Java, Python, Ruby, Erlang, Perl, PHP, C#, etc.,
(j.) Cross-platform: UNIX, Linux, Windows, Mac OS, Sun Solaris…
(k.) Simple library API (Application Programming Interface),
(l.) Simple installation of FUSE,
(m.) Usable by non privileged users,
(n.) very stable over time,
(o.) Highly portable and secure systems can be developed with FUSE.

There is also a potential disadvantage of user space file systems that they degrade performance of the system as compared to the kernel level implementation. There are additional context switches and memory copies overhead.

The request of the real information of a file will be passed from layer to layer through drivers and interfaces until request handling program written by user in the user space.

The Fig. 3 shows the data flow used by FUSE to access remote data. FUSE contains three modules: FUSE kernel module, libfuse module, User program module. In user space, users should implement the file system which is encapsulated by the Libfuse library. Libfuse provide the support to the main file system framework, encapsulation for -user implemented file system code, handling -mount, communication with operating system module through character device.

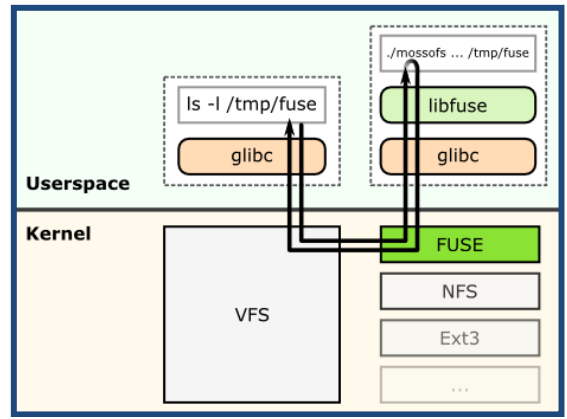

Figure 3. FUSE Architecture

Work flow is described as below:
(a) File system request by the application to the kernel
(b) VFS issues the request and send it to FUSE Module within the kernel
(c) FUSE module receive it and forward it to the FUSE file system
(d) The FUSE file system receives it and gives response to the FUSE module back
(e) FUSE module receives the response from the file system and send it to the VFS kernel module
(f) Finally the response is given to the concern Application

The kernel module of FUSE has implemented the VFS interface which is used for FUSE file driver module registration, the virtual device driver of FUSE, providing maintenance of supper block, dentry, inode. FUSE kernel will receive the VFS's requests and pass them to libfuse. Libfuse will pass them to our user program interface to actually do the job (Fig. 3).

It is well observed that the FUSE file system approach is not good because of the performance issues. It is also a new idea.

## IV. KERNEL LEVEL PROCESS FILE SYSTEM

To develop a file system in the operating system kernel means forgoing the simplicity of development that a user process provides. This increases development complexity because kernel level programming requires specialist knowledge of the specific operating system being used. When the file system resides in the kernel, a tight coupling exists between the file system and the kernel level services that it uses. This coupling reduces the ability of the file system developer to port the file system to another operating system.

Developing the file system from scratch inside the kernel allows the file system developer greater freedom in the implementation process. Gaining experience with the internals of the kernel requires time and considerable knowledge of the underlying operating system structure. Developing a file system from first principles does not utilize any of the development previously done. Redeveloping all of the file system functionality in this way does not make any sense.

Examples of file systems developed in the kernel include the NFS. Fig. 4 illustrates how a typical user level process utilizes the file system operations in the kernel. It requires two calls, (1) for the request and (2) for the response.

For the majority of file systems developed in the kernel (in UNIX based operating systems) the system calls that are made to the kernel are routed through the Virtual File System layer (VFS). The VFS layer allows the kernel to provide access to different file systems through a common interface. The kernel provides a data structure called a vnode to represent an open file or socket without revealing the underlying file system implementation. All operations performed on vnodes are the same regardless of the underlying file system implementation.

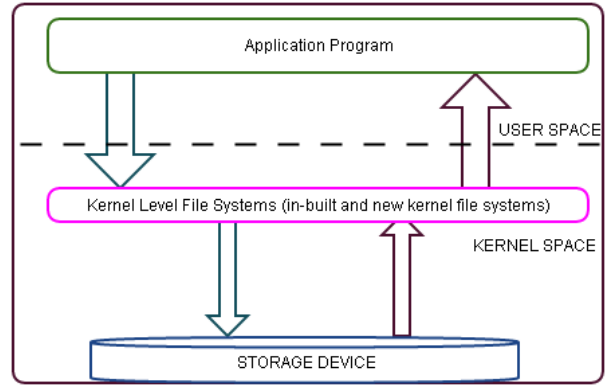

Figure 4. File System implemented in the kernel.

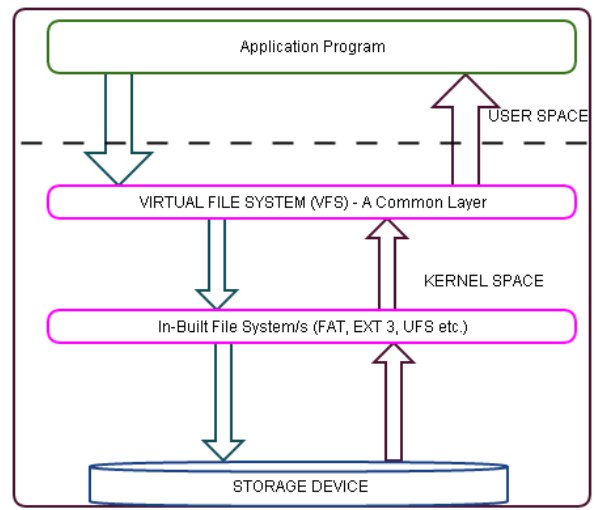

Figure 5. VFS Supports Multiple File Systems in the kernel.

Fig. 5 shows how file system operations are routed through the VFS vnode to the underlying file system ext2. The NFS file system is also present to illustrate how a user could access remote files. This serves to illustrate how multiple file systems are catered for in the same kernel.

## V. STACKABLE LAYER KERNEL LEVEL PROCESS FILE SYSTEM

Significant work has been done using stackable layers to leverage existing functionality provided by file systems implemented in the kernel. The extensible file systems in Spring, Lofs, Rot13fs and Usenetfs are examples of file systems that use stackable layers and are discussed in "A Stackable File System Interface For Linux". The Fiscus Replicated File System describes how stackable layers provide replication of files. An implementation of a cryptographic file system in Linux, Cryptfs demonstrates how stackable layers can be used to create a useful file system by leveraging the existing file system functionality. Stackable layers use the VFS interface and vnodes to layer functional operations one on top of the other.

The process of developing file system functionality in the kernel is difficult due to the constraints that the kernel imposes. It is preferable to reuse existing code whenever possible as it has usually been thoroughly tested and is generally stable. The main idea behind stackable layers is to reuse existing functionality by layering new functionality on top of it. Developers can reap the benefits of previous work and concentrate on the problems associated with their required functionality.

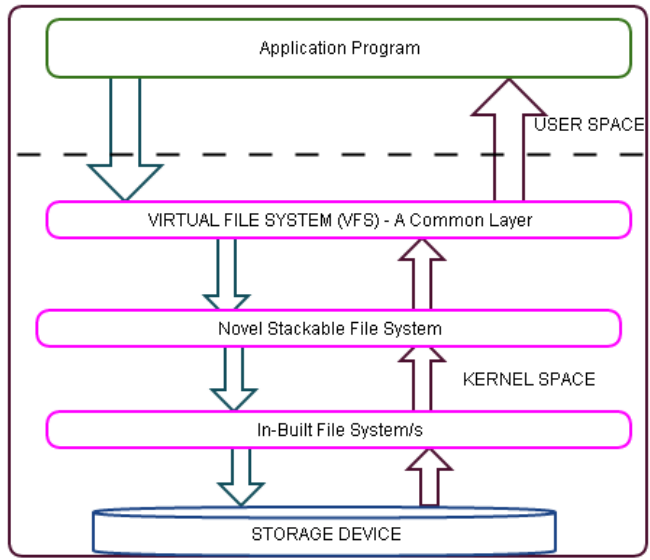

Figure 6. New File System functionality added to the Kernel as a Stackable Layer.

The Fig. 6 shows how a stackable layer is used inside the kernel to utilize existing functionality.

## CONCLUSION

The article has elaborated some of the fundamental file system design approaches. It is concluded that the kernel based file systems are more efficient as compared to the user based file system. But the kernel based file systems may harm the complete system if they are not designed properly and the complete system may face huge consequences because of this. The article has given the great insight for the file system design approaches.